\documentclass[manuscript]{aastex}

\usepackage{multirow}

\newcommand{\al}{{$\alpha$}}

\shorttitle{Constraining Magnetic Field Extrapolation}
\shortauthors{Conlon et al.}

\begin{document}

\title{Constraining 3D Magnetic Field Extrapolations Using The Twin Perspectives of \emph{STEREO}}

\author{Paul A. Conlon and Peter T. Gallagher}
\affil{Astrophysics Research Group, School of Physics, 
   	Trinity College Dublin, Dublin 2, Ireland}

\begin{abstract}
The 3D magnetic topology of a solar active region (NOAA 10956) was reconstructed using a linear force-free field extrapolation constrained using the twin perspectives of \emph{STEREO}. A set of coronal field configurations was initially generated from extrapolations of the photospheric magnetic field observed by the Michelson Doppler Imager (MDI) on \emph{SOHO}. Using an EUV intensity-based cost function, the extrapolated field lines that were most consistent with 171~\AA\ passband images from the Extreme UltraViolet Imager (EUVI) on \emph{STEREO} were identified. This facilitated quantitative constraints to be placed on the twist ($\alpha$) of the extrapolated field lines, where $\nabla \times {\bf B} = \alpha {\bf B}$. Using the constrained values of $\alpha$, the evolution in time of twist, connectivity, and magnetic energy were then studied. A flux emergence event was found to result in significant changes in the magnetic topology and total magnetic energy of the region.

\end{abstract}


\section{Introduction}

Extreme solar events such as solar flares and coronal mass ejections (CMEs) originate in the corona of active regions. As the corona is a low-$\beta$ plasma, the magnetic field rather than the gas pressure governs its structure and evolution \citep{gary:2001}. This enables force-free field extrapolations of the photospheric field to be used to study the detailed topology of the solar corona \citep{longcope:2009}. Of particular interest to space weather applications is how the three-dimensional (3D) field topology of active regions governs the production of extreme solar events. Unfortunately, previous attempts to reconstruct the 3D topology of active region fields have been limited by observational issues, such as incomplete measurements of the photospheric vector field, or theoretical issues such as ill-constrained extrapolation algorithms (e.g., \citeauthor{derosa:2009} \citeyear{derosa:2009} and references within). With the launch in December 2006 of the \emph{Solar TErrestrial RElations Observatory (STEREO)} satellites, the observational limitations on reconstructing the 3D topology of active regions has largely been overcome, thus opening a new era in the study of the ever-changing solar corona.

Observations of the coronal magnetic field are difficult to obtain due to the tenuous nature of the corona and the absence of strong magnetically sensitive emission lines. Estimates of the strength of the coronal magnetic field are possible using radio observations, however, this remains a difficult problem requiring simultaneous knowledge of both the temperature and density of the solar atmosphere~\citep{brosius:2006}. In order to overcome these issues, modellers have developed a variety of theoretical techniques to examine the topology of the coronal magnetic field~\citep{gary:1989,mikic:1990}. These methods require observations of the photospheric or chromospheric magnetic field as boundary conditions for the extrapolations. Well constrained extrapolations are normally produced using photospheric vector magnetograms, although \citet{metcalf:1995,metcalf:2002} have developed methods that make use of chromospheric vector magnetograms. As the chromosphere is in a force-free state, chromospheric vector magnetograms are ideally suited as boundary conditions in force-free extrapolation methods. However, these observations have a limited field-of-view and spatial and temporal resolution. As such, photospheric magnetograms are commonly used as boundary conditions for magnetic field extrapolations.

The extrapolation of the photospheric field into the corona requires that a number of assumptions be made. Firstly, it is assumed that the coronal magnetic field remains stable for the duration of the observation. Therefore, a static magnetic field model is normally used. In addition, it is reasonable to assume that the field is in a force-free configuration. This results in a vanishing Lorentz force, $\mathbf{j}\times\mathbf{B} =  0$, where $\mathbf{j}$ is the current density, $\mathbf{B}$ is the magnetic field. The force-free condition requires that the current density remains parallel to the magnetic field, $\mu_0\mathbf{j} = \mathbf{\alpha} \mathbf{B}$, and therefore $\mathbf{\nabla} \times \mathbf{B} =\mathbf{\alpha} \mathbf{B}$, where $\mathbf{\alpha}$ is the amount of twist in the field and is some times referred to as the torsion function. In the case of $\mathbf{\alpha} = \alpha(\mathbf{r})$, the governing equations becomes $\mathbf{\nabla} \times \mathbf{B} =  \alpha(\mathbf{r}) \mathbf{B}$ and $\mathbf{B}\cdot\nabla\alpha(\mathbf{r})=0$. This non-linear force free (NLFF) field is the generic solution to the force-free field condition. Complete knowledge of the photospheric magnetic field allows for the inversion of the preceding equations and the calculation of $\alpha$ at each point on the photosphere. As such, vector magnetograms are required to calculate the NLFF field in the corona~(\citeauthor{Schrijver:2006} \citeyear{Schrijver:2006}). While NLFF methods offer the possibility of increased accuracy, they are computationally intensive. Additionally, NLFF extrapolation suffer from several other sources of error, such as the limited field-of-view, non-force-free nature of the photospheric field, the intrinsic 180-degree ambiguity, and low signal to noise of the transverse component of vector magnetograms~\citep{derosa:2009}. Forcing $\mathbf{\alpha} = 0$ results in the potential or current-free solution, which is a first order approximation to the coronal magnetic topology. The potential field solution is the first of two subclasses of solutions to the force-free field. Given the reduced complexity of the potential solution, the only boundary condition necessary to form a solvable set of equation are measurements of the line-of-sight (LOS) magnetic field~\citep{gary:1989}. The second and final subclass of solution to the force-free field is the linear force-free (LFF) field, where $\mathbf{\alpha}$ = constant and $\mathbf{\nabla} \times \mathbf{B}=\alpha \mathbf{B}$. Assuming that \al{} is constant for all field lines allows for the formation of a closed set of equations with only knowledge of the LOS magnetic field, provided \al{} is predefined~\citep{alissandrakis:1981}. 

A major difficulty in studying the extrapolated field line distribution is deciding on what is the most appropriate way to constrain $\alpha$. \citet{leka:1999} and more recently \citet{burnette:2004} compared three methods and applied them to a single active region. The first method considered was based on the best fit value of $\alpha$ from a LFF extrapolation of a longitudinal magnetogram that best matched the horizontal field of the corresponding vector magnetogram~\citep{pevtsov:1995}. Secondly, the value of $\alpha$ was chosen from the mean of the distribution of local $\alpha_z = \alpha(x,y)$~\citep{pevtsov:1994}. In the final method, the value of $\alpha= J_z/B_z$ was inferred from the slope of a least-squares linear fit to the distribution of local $J_z(x, y)$ versus $B_z(x, y)$.  \citet{burnette:2004} conducted a detailed study of these three methods, concluding that they give statistically consistent values of $\alpha$. As the methods described above require the availability of high quality vector magnetograms, they are not always an available option for constraining the values of $\alpha$.

EUV and X-ray images of emitting coronal structures, mainly loops, can also be used to constrain the choice of \al{} in magnetic field extrapolations. Initially, loop tracing methods were used to provide a first guess at the structure of coronal loops and through the application of numerical methods the 3D structure found~\citep{aschwanden:1999}.  Other authors have developed methods that minimize the separation between the field lines calculated from observations and extrapolations as a constraint on the choice of \al{}~\citep{lim:2007}. \citet{wiegelmann:2005},~\citet{carcedo:2003}, and ~\citet{aschwanden:2008a} developed related techniques that examine the EUV emission of coronal loop structures and retrieve their two dimensional structure using a cost function method. \citet{wiegelmann:2005} developed a cost function grading method to evaluate the agreement of calculated field lines and the observed emission of coronal loops. This enabled them to gauge the optimum value of $\alpha$ for a small number of LFF extrapolations. The cost function method was found to allow for the selection of the optimum value of $\alpha$ used in the LFF extrapolations, however, the number of possible values of $\alpha$ examined was insufficient to allow for a detailed study of the active region. Most recently, \citet{aschwanden:2008,aschwanden:2009} expanded on the methods discussed above with the use of the \emph{STEREO} spacecraft to study the structure and oscillations of coronal loops. 

Once the value of the magnetic twist ($\alpha$) in a region has been well constrained, an accurate estimate of the magnetic energy or more importantly, the free magnetic energy, can be estimated. The magnetic energy budget of active regions can be calculated using volume-integral and surface-integral methods such as the magnetic Virial Theorem~\citep{emslie:2004}. The free-energy is calculated as the difference between the magnetic energy of the NLFF field and that of a lower bound such as the potential (current-free) field. These methods have shown the free energy in active region's to be in the range of $10^{32}-10^{33}$~ergs depending on the size of the region~\citep{gary:1989,regnier:2007,metcalf:1995, metcalf:2005}. The magnetic Virial Theorem assumes the measured field is a force-free field and is only possible if the measured data is force-free consistent. Additionally, knowing the topology of the active region at instances during its evolution, changes in the regions connectivity can be examined. Studies of the changing connectivity of active regions using a combination of active region segmentation techniques and magnetic field extrapolations have been achieved~\citep{longcope:2001,longcope:2007,longcope:2009}. \citet{longcope:2009} concluded that comparison of the changing connectivity in active region over time could provide insight into energetics and reconnection in the coronal fields.

The methods described in this paper allow for the 3D structure of the magnetic field of an active region to be recovered and a detailed analysis of its evolution to be performed using \emph{SOHO} Michelson Doppler Imager (MDI,~\citeauthor{scherrer:1995} \citeyear{scherrer:1995}) LOS magnetograms and \emph{STEREO} Extreme UltraViolet Imager (EUVI, \citeauthor{howard:2000}~\citeyear{howard:2000}) images. The observation and the transformation of co-ordinates between \emph{STEREO} and \emph{SOHO} reference frames are outlined in Section~\ref{observation}, while details of the extrapolation method and the cost function are then given in Section~\ref{method}. Our results, relating to twist, connectivity and magnetic energy are presented in Section~\ref{results}, and our conclusions are given in Section~\ref{conclusion}.

\section{Observations}
\label{observation}

NOAA 10956 produced a \emph{GOES} class B9 flare on 2007 May 19. Associated with this flare was a CME and a globally propagating disturbance within the solar corona~\citep{long:2008}. Extrapolations of the region's coronal magnetic field were constrained at the lower boundary with LOS magnetic field measurements obtained by MDI on \emph{SOHO}. MDI images the Sun on a $1024\times1024$ pixel$^2$ CCD camera through a series of increasingly narrow filters~\citep{scherrer:1995}. A pair of tunable Michelson interferometers enable MDI to record filtergrams with a FWHM bandwidth of 94 m\AA{}. 96-minute cadence magnetograms of the solar disk with a pixel size of $\sim 2''$ are used in this work. 

Images from EUVI, part of the Solar Earth Connection Coronal and Heliospheric Investigation (SECCHI, \citeauthor{howard:2000}~\citeyear{howard:2000}) suite of instruments on board both \emph{STEREO} (\emph{A})\emph{head} and (\emph{B})\emph{ehind}, were used in order to constrain the choice of $\alpha$ within the LFF extrapolation of the corona. In particular, 171~\AA{} images from EUVI were used as this passband has fewer contributions from diffuse higher temperature emission lines compared to 195~\AA{} images~\citep{phillips:2005}. This results in more defined loops which are easier to compare by the methods described in Section~\ref{method}. Each \emph{STEREO} image was co-aligned with a common \emph{SOHO} MDI image using a transformation method based on the work of \citet{aschwanden:2008}, so as to allow the transformation of points between the \emph{STEREO A}, \emph{B} and \emph{SOHO} reference frames. 

\section{Methods}
\label{method}

\begin{figure}[!b]
%\centerline{\includegraphics[width = 0.75\linewidth]{line_profile.eps}}
\centerline{\includegraphics[width = 0.5\linewidth]{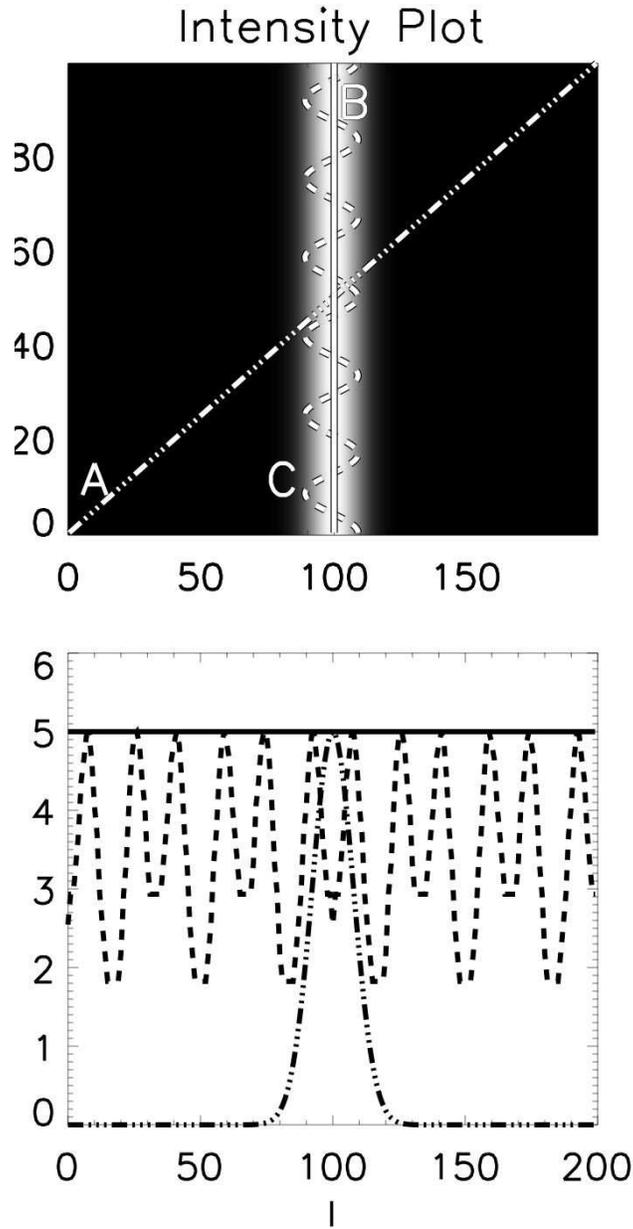}}
\caption{Field lines overlaid on emission with a Gaussian profile. Top: Bright Gaussian ridge, with three overlaid field lines `A', `B', and `C'. Field line `A' crosses the ridge, `B' rest along the ridge, and `C' oscillates along the ridge. Bottom: Intensity as measured along each field line.}
\label{cost}
\end{figure}

Given the limited reliability of vector magnetograms and the errors associated with their use in NLFF methods, a LFF methods is used in this study to recover the coronal field as accurately as possible. Assuming a force-free field, $\mathbf{j}\times \mathbf{B} = \mathbf{0}$, a LFF field is governed by:
\begin{equation}
\mathbf{\nabla} \times \mathbf{B}=\alpha \mathbf{B},
\end{equation}
where $\alpha =$ constant. For LFF field extrapolations, $\alpha$ is a predefined free parameter. Small changes in $\alpha$ result in large changes to the extrapolated field. Comparing extrapolated fieldlines with EUV observations provides a method of restricting $\alpha$ and improving the accuracy of our results. 

It is possible to systematically determine the value of $\alpha$ that provides the best possible comparison with the EUV observations using the following steps:
\begin{enumerate}
\item Compute the linear force-free field.
\item Calculate a large number of possible magnetic field lines.
\item Project field lines onto EUV observation.
\item Calculate the cost function along each field line.
\item Repeat for each additional value of $\alpha$.
\item Compare the cost of the best matched field lines for each $\alpha$.
\end{enumerate}

In order to do this accurately it was first important to define what a \emph{good fieldline} is. Figure~\ref{cost} shows the path and resulting intensity for three different ideal field lines. Field line `A' crosses the bright emission only once and has a small spike in emission along its length. Field line `B' rests along the field line and has a smooth bright intensity profile. Field line `C' oscillates on and off the bright emission and the resulting profile has a series of spikes in the intensity. With this in mind a number of methods have been developed to systematically extract loop profiles from EUV image of the solar corona~\citep{carcedo:2003,wiegelmann:2005}. Including this information in the definition of a cost function used to constrain the choice of \al{} removes the intermediate step of extracting coronal loop profiles. For the purposes of this research, modifications to the \citet{wiegelmann:2005} method were studied to find the most computationally efficient and best performing cost function. A small sample of the various cost functions studied are:
\begin{eqnarray*}
C_{W} = & \int \nabla I(l) dl  \Big/ \Big(l(\int I(l) dl)^2\Big),\\
C_{EW} = & \int \nabla I(l) dl  \Big/\Big( l\int I(l) dl \Big),\\
C_{B} = & 1/\int I(l) dl,
\end{eqnarray*}
where $I(l)$ is the intensity of emission along a loop and $l$ is the loop length, $C_{W}$ and $C_{EW}$ are the standard and equal weight cost functions as defined by~\citet{wiegelmann:2005}, and $C_{B}$ is a modified cost function developed to select the brightest field lines. 
\begin{figure*}[!t]
\centerline{\includegraphics[width = \linewidth]{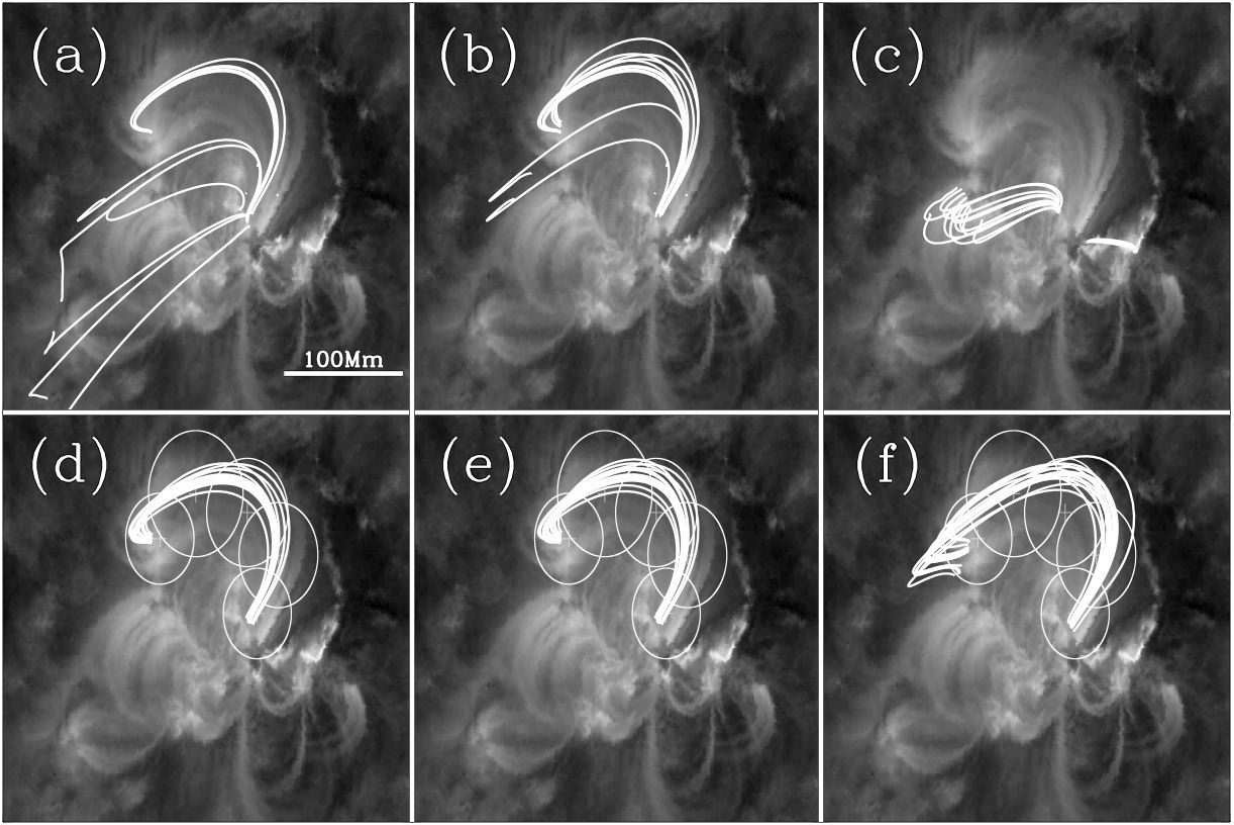}}
\caption{Field lines that best represent the observed emission as calculated by the selection algorithm for $\alpha= 6.25\times10^{-3}$~Mm$^{-1}$. (\emph{a}) $C_{W}$, (\emph{b}) $C_{EW}$ and (\emph{c}) $C_{B}$ cost functions without field line path restrictions. (\emph{d, e, f}) same with field line path restriction as outlined with overlaid circles.}
\label{cost_compare}
\end{figure*}

\begin{figure}[!t]
%\centerline{\includegraphics[width = 0.75\linewidth]{compare_a_B_fieldlines.eps}}
\centerline{\includegraphics[width = 0.5\linewidth]{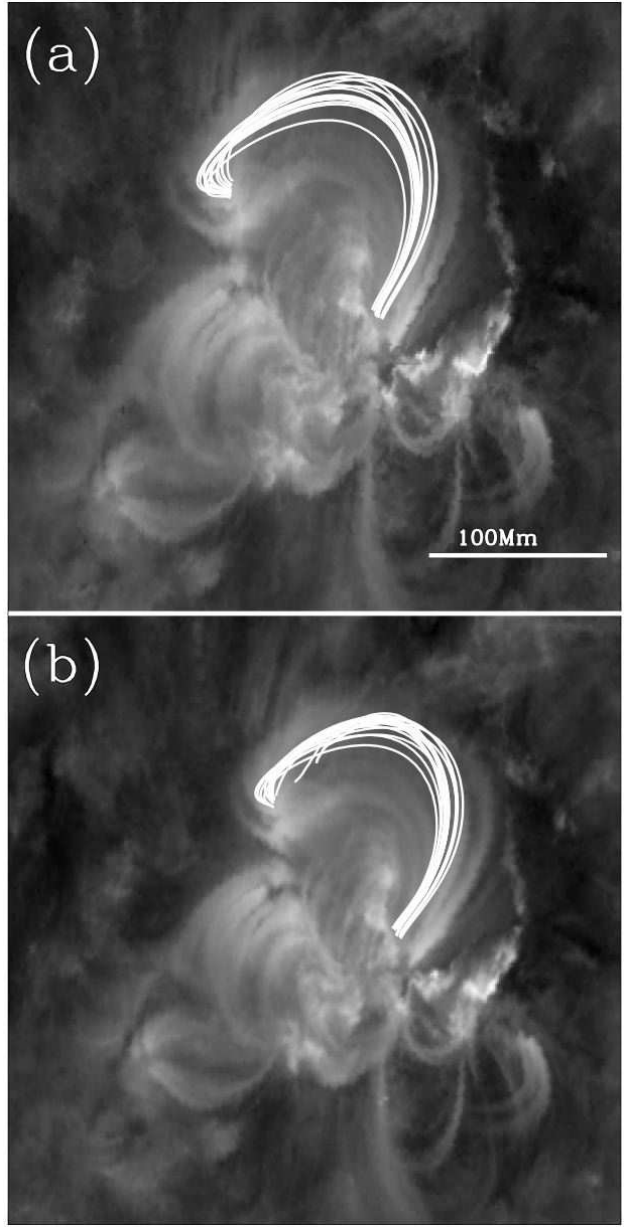}}
\caption{(\emph{Top}) 15 field lines with the smallest cost as selected using the $C_{EW}$ cost function with \emph{STEREO A} EUVI images of NOAA 10956. (\emph{Bottom}) Same but for \emph{STEREO B} images,  for $\alpha$ = $8.7\times10^{-3}$~Mm$^{-1}$.}
\label{A_B_compare}
\end{figure}

\begin{figure}[!t]
\centerline{\includegraphics[width = \linewidth]{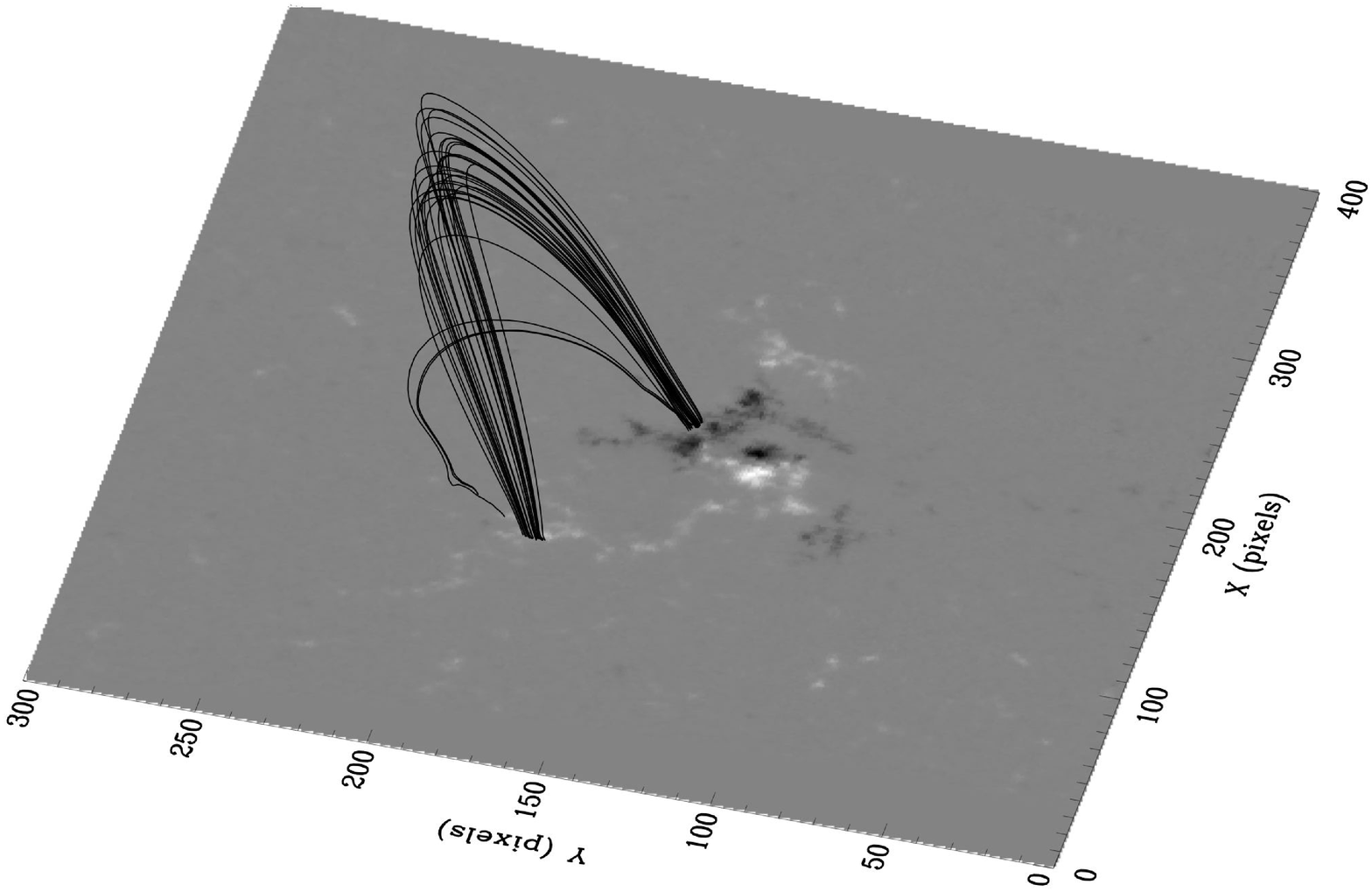}}
\caption{3D view of field lines shown in Figure~\ref{A_B_compare}.}
\label{3D_plot}
\end{figure}

Using the LFF method from \citet{alissandrakis:1981}, field lines were extrapolated starting from a user-defined foot-point. Foot-points were selected from EUV observations of the coronal loops in question. Field lines were then transformed to the desired perspective of each \emph{STEREO} spacecraft, and the emission along each field line and associated cost calculated. Figure~\ref{cost_compare} (\emph{Top}) show the ability of the three cost functions to recover the geometry of the active region loop. As can be seen, the $C_{EW}$ and $C_{W}$ cost functions perform best, both however include unrealistic field lines. These are due to the infinite number of possible field line paths leaving any sub-region of the active region. Given the tenuous nature of the coronal loops we wish to study, field lines that cross the core of the region and other high emission locations will have unacceptable paths and relatively low costs.

In order to overcome this, fieldlines were forced to emerge from user defined foot-points and pass through a number of guide circles. These user defined guide-circle domains are highlighted in Figure~\ref{cost_compare} (\emph{Bottom}). Figure~\ref{cost_compare} (\emph{Bottom}) shows the ability of the three cost functions to recover the geometry of the active region loop with the inclusion of these guiding circles. This can be seen to greatly increase the ability of each cost functions to recover the coronal geometry. The $C_{EW}$ cost function was used in this work to select the value of $\alpha$ for the LFF extrapolations due to its ability to select more valid field lines under both conditions. The user-defined foot-points and guide circles stop the algorithm from selecting loops and other bright features outside the region of interest. Large guide circles, with radii less that a typical loop half-length, were chosen to ensure that difference user selections would result in similar features being used to constrain the extrapolations. It may, in the future, be possible to automate the selection of coronal loops based on EUV and/or X-ray emission, but this will require the application of an accurate and robust coronal loop detection algorithm (e.g \citeauthor{aschwanden:2008a}, \citeyear{aschwanden:2008a}).

Using the $C_{EW}$ cost function it is possible to compare the cost function for various values of $\alpha$ and determine the value that most closely resembles the observed field. More than 5,000 field lines are plotted for each value of $\alpha$ and the cost along each is calculated. In order to enhance the sensitivity of the method to changes in the values of $\alpha$ only the total cost of the 15 best field lines are compared. As shown in Figure~\ref{cost_compare}, this is a sufficient number of field lines to accurately recover the topology of the region. The contrast of the cost for each value of $\alpha$ was increased further by normalizing the returned values and setting the cost to $1$ for values of $\alpha$ where no field lines pass through the guiding path and/or end at the user-defined foot-point.

Figure~\ref{A_B_compare} shows the best field lines as selected by the method for \emph{STEREO} EUVI images from both the \emph{A} and \emph{B} spacecrafts. The field lines returned by the method are not the same for each spacecraft. However, the general geometry defined by the field lines is very similar. Figure~\ref{3D_plot} shows the field lines from both perspectives plotted together in 3D.

An estimate of the energy of the LFF field in excess of the potential field can be obtained from:
\begin{eqnarray}
E_{Diff} &=& E_{LFF} - E_{Pot},
\end{eqnarray}
where $E_{LFF}$ is the magnetic energy of the LFF field extrapolation, 
\begin{eqnarray}
E_{LFF} &=& \int (\mathbf{B}_{LFF}^2/8\pi) dV,
\end{eqnarray}
and $E_{pot}$ is the energy of the potential field extrapolation,
\begin{eqnarray}
E_{Pot} &=& \int (\mathbf{B}_{Pot}^2/8\pi) dV,
\end{eqnarray}
where $dV$ is a volume within the computational domain, and $E_{Pot}$ and $E_{LFF}$ are the magnetic energy of the potential field and LFF field respectively.

As a precautionary warning, it should be noted that LFF extrapolations have a maximum value of $\alpha$ that can be studied before oscillatory signals from the complex solution of the governing set of equation begin to dominate the systems~\citep{gary:1989}. We have investigated this effect for the data set studied here and found that for $\alpha$ values up to $0.03$~Mm$^{-1}$ the oscillatory signal from the complex solutions is negligible up to a height of $100$~Mm, the maximum height of loops studied here.

\section{Results}
\label{results}

% change in alpha
\begin{figure}[!b]
\centerline{\includegraphics[width = 0.75\linewidth]{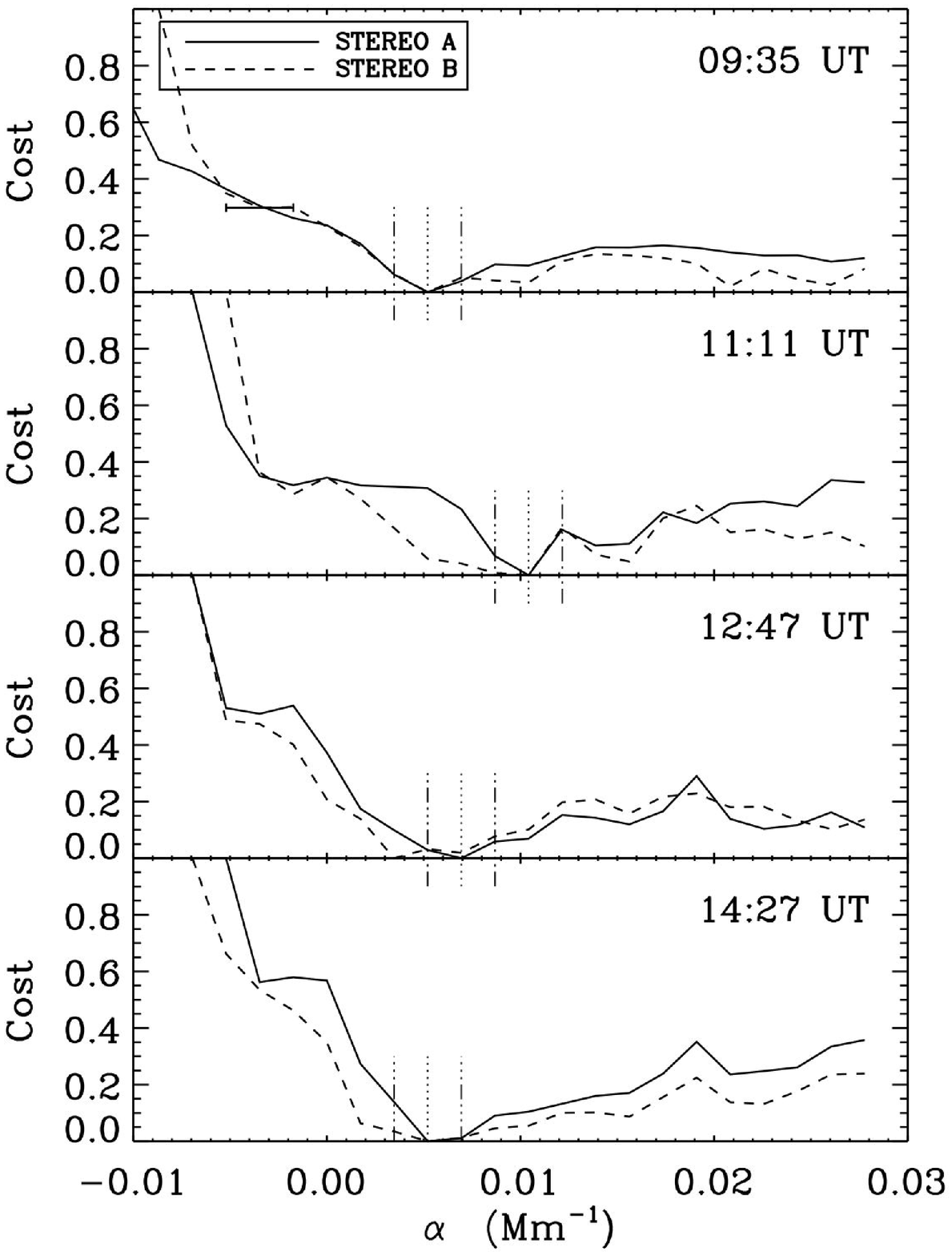}}
\caption{Top to bottom: Cost for a range of $\alpha$ values for MDI magnetograms obtained at 09:35~UT, 11:11~UT, 12:47~UT, and 14:27~UT. Solid and dashed lines represent comparisons with \emph{STEREO A} and \emph{B} EUVI observations respectively. Error bars in top panel represent the associate error in all the values of $\alpha$ due to the step-size used in the calculations. Minimum in the cost function and the associated error are highlighted by the vertical dotted lines and dashed-dotted lines respectively. The active region produced a B9 flare that began at $\sim$12:34~UT, peaked at $\sim$13:02~UT, and ended at $\sim$13:19~UT.}
\label{alpha_time}
\end{figure}

Using the method described in Section~\ref{method}, an initial investigation of the cost function behavior for a range of $\alpha$ values from $-0.3$~Mm$^{-1}$ to $0.3$~Mm$^{-1}$ had established that a minimum in the cost function existed and had a positive value of $\alpha$. To verify this a more detailed analysis of the cost function was conducted for twenty three values of $\alpha$ within the range of $-0.11$~Mm$^{-1}$ to $0.28$~Mm$^{-1}$. Figure~\ref{alpha_time} shows the evolution of the cost function for these values of $\alpha$ at the times studied. At all times there is a strong correlation between both the \emph{STEREO A} and \emph{B} cost functions. At 09:35~UT there is a distinct minimum in the cost at $5\times10^{-3}$~Mm$^{-1}$. The cost for the \emph{STEREO B} observations was seen to decrease for larger values of \al{}, however as the \emph{STEREO A} cost does not follow this trend it is assumed false. At 11:11~UT the minimum has increase to around $1\times10^{-2}$~Mm$^{-1}$, the relatively high cost returned for \al{} =   $5\times10^{-3}$~Mm$^{-1}$ reinforces this increase. From 12:47~UT onwards the minimum cost in \al{} or twist of the coronal loop analyzed is seen to decrease to levels seen at 09:35~UT.

% Change in B
\begin{figure*}[t]
\centerline{\includegraphics[width = \linewidth]{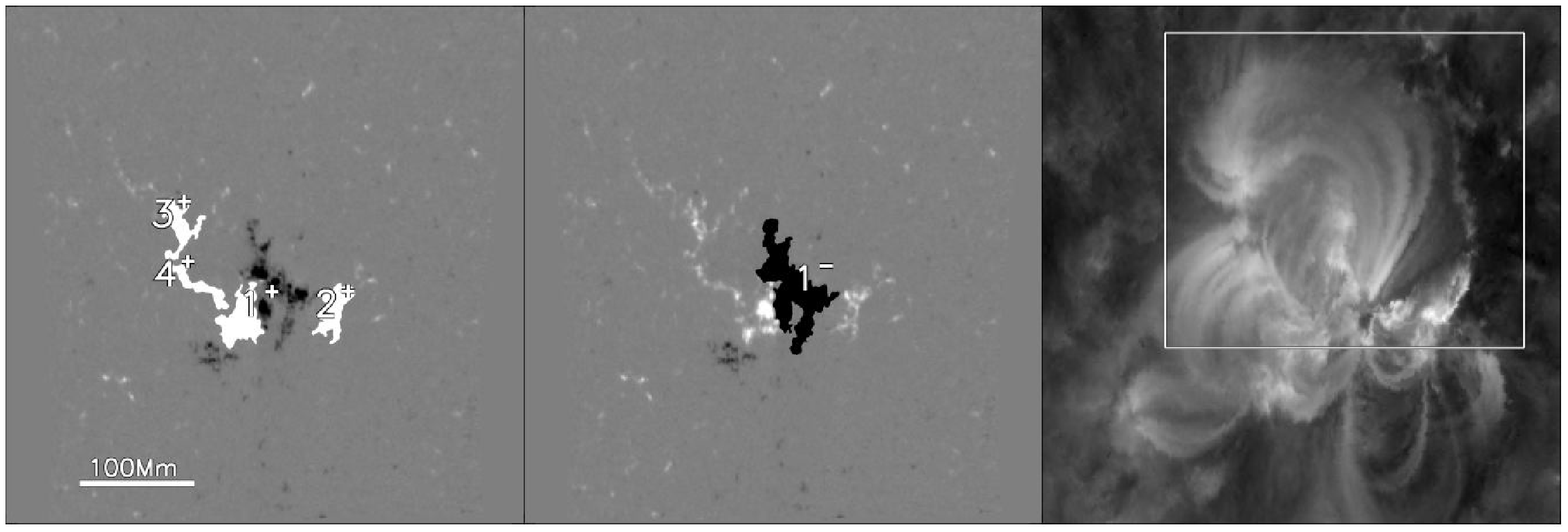}}
\caption{Sub-regions of NOAA 10956. Left panel: Positive regions identified by thresholding, ranked in order of area. Center panel: Same for negative regions. Right panel: Corresponding \emph{STEREO}/EUVI image. The smaller volume centered on the observed loop used for the energy calculations is illustrated by the box shown.}
\label{regions}
\end{figure*}% What do we see

The evolution of magnetic flux within each sub-region was examined in order to more fully understand these changes in \al{}. Each magnetogram image of NOAA 10956 was thresholded at the $\pm 100$G level to identify the main regions of magnetic flux in the global structure. Figure~\ref{regions} shows the main positive and negative sub-regions within the active region, ranked by area from largest to smallest. From the corresponding \emph{STEREO}/EUVI image it is clear that the large loop structure studied here is associated with field lines connecting regions 1$^-$ and 3$^+$.  Table~\ref{mag} summaries the changing magnetic flux within the region (some negative regions are excluded due to changes in ranking during the time studied). 

%\clearpage
\begin{deluxetable}{c c c c c c }
%\rotate
\tabletypesize{\scriptsize}
\tablecaption{Evolution of the magnetic flux in the major sub-regions contained within NOAA 10956. Results are reported in $10^{20}$~Mx.\label{mag}}
\tablewidth{0pt}
\tablehead{
\colhead{Region} & \colhead{09:35~UT} & \colhead{11:11~UT} & \colhead{12:47~UT} & \colhead{14:27~UT}}
\startdata
1$^{-}$ &77.8 & 76.5 & 76.5 & 76.9\\ 
1$^{+}$ & 32.5 & 33.1 & 33.7 & 34.5\\ 
%\cline{2-6}
%\tableline
2$^{+}$ & 12.5 & 11.4 & 11.7 &11.9\\
%\tableline
3$^{+}$ & 7.2 & 7.1 & 6.8 & 8.6\\  
%\tableline
4$^{+}$ & 3.3 & 3.4 & 3.7 & 2.1\\ 
\enddata
\end{deluxetable}

As can be seen there was a significant changes in the magnetic flux in the region during the time analyzed.  From 09:35~UT to 11:11~UT, the magnetic flux in region 1$^+$ increase by $0.6\times10^{20}$~Mx, while in regions 1$^{-}$ and 2$^{+}$ the magnetic flux decreased by $1.3\times10^{20}$~Mx and $1.1\times10^{20}$~Mx respectively. During this time the other sub-regions saw little or no changes in their magnetic flux. From 11:11~UT to 12:47~UT, the magnetic flux in regions 1$^{+}$ and 2$^{+}$ increased by $0.6\times10^{20}$~Mx and $0.3\times10^{20}$~Mx respectively. While the magnetic flux in region 3$^{+}$ decreased by $0.3\times10^{20}$~Mx. From 12:47~UT to 14:27~UT, the magnetic flux in regions 1$^{+}$, 1$^{-}$, and 3$^{+}$ increased by $0.8\times10^{20}$~Mx, $0.4\times10^{20}$Mx, and $1.8\times10^{20}$~Mx respectively.

% change in connectivity.
Knowing the values of $\alpha$, it is then possible to study changes in the connectivity within the region. As region $1^-$ is the common foot-point for most of the loops in the active region it was selected as the source for all the extrapolated field lines used to examine the region's changing connectivity. Table~\ref{connectivity} outlines the changing connectivity for the time studied. From 09:35~UT to 11:11~UT the major changes in connectivity are associated with the transfer of flux from a closed to open configuration and a 1.9\% increase in flux connecting region 1$^-$ to region 4$^+$. From 11:11~UT to 12:47~UT, there is a decrease in the amount of open flux and an increase in the connectivity from region 1$^-$ to regions 1$^+$, 2$^+$, and 3$^+$ by 4.8\%, 0.8\%, and 2.3\% respectively. Additionally, the amount of flux joining region 1$^-$ to region 4$^+$ decreases by 3\%. From 12:47~UT to 14:27~UT the amount of open flux continues to decrease and there is no flux left connecting regions 1$^-$ to 4$^+$. The amount of flux connecting region 1$^-$ to region 1$^+$ increases by an additional 9\%. There are marginal changes in the flux connecting 1$^-$ to 2$^+$ and 3$^+$. 

%\clearpage
\begin{deluxetable}{c c c c c}
%\rotate
\tabletypesize{\scriptsize}
\tablecaption{Connectivity matrix for field lines starting at region 1$^-$, with threshold of $\pm 50$~G. All results are a percentage of total flux leaving region 1$^-$.\label{connectivity}}
\tablewidth{0pt}
\tablehead{
\colhead{1$^{-}$} & \colhead{09:35~UT} &\colhead{11:11~UT} & \colhead{12:47~UT} & \colhead{14:27~UT}}
\startdata
Open &13.9 & 20.6 &15.9 &12.9  \\      
Closed &31.9 &24.3 &23.8 &23.6\\
1$^{+}$ & 24.5 & 25.1 & 29.8 & 38.9\\
2$^{+}$ & 11.1 &10.8 &11.6 & 11.2\\
3$^{+}$ & 9.8 & 9.7 & 12.0 & 12.3\\ 
4$^{+}$ & 7.4 & 9.3 & 6.2 & 0.00
\enddata
\end{deluxetable}

% change in Energy

Following from the investigation of the changing magnetic flux within the region and the significant changes in connectivity, an investigation was undertaken into the energy stored in the magnetic field. Table~\ref{energy} shows the calculated magnetic energy for two volumes: one centered on the loop structure connecting regions 1$^-$ and 3$^+$ (see boxed region outlined in Figure~\ref{regions}) and the other being the full computational volume. An estimate of the error in the energy calculations is provided by calculating the energy of the LFF field for $\alpha_{min} - \Delta_{\alpha}$ where $\Delta_{\alpha}=0.17$\footnote{The step size of $\alpha$ during the cost function analysis is $\Delta_{\alpha}=0.17$} and $\alpha_{min}$ is the calculated $\alpha$ value of the loop observed. From this it is clear that there is a significant buildup in energy within both volumes from 09:35~UT to 11:11~UT. At 12:47~UT, the difference in magnetic energy between the potential and LFF field has significantly decreased and is only marginally higher than that at 09:35~UT, while at 14:27~UT this difference in magnetic energy is lower than at 09:35~UT.

\begin{deluxetable}{ c c c c c c }
\tabletypesize{\scriptsize}
\tablecaption{Changes in the magnetic energy contained within different volumes in NOAA 10956. The small region is a box centered on the the loop structure connecting regions 1$^-$ and 3$^+$. The full volume in the whole computation domain. All values for energy are reports in units of $10^{32}$~ergs.\label{energy}. Associated errors in the calculation of the LFF energy and the energy difference are on average $\pm0.01$ ($\pm0.03$), and a maximum of $\pm0.03$ ($\pm0.05$) at 11:11~UT for the small region (full volume).}
\tablewidth{0pt}
\tablehead{
\colhead{Size} &\colhead{Parameter} &\colhead{09:35~UT} & \colhead{11:11~UT} & \colhead{12:47~UT} & \colhead{14:27~UT}
}
\startdata
Small  Region &Pot & $4.46$&$4.44$&$4.44$&$4.48$ \\     
&LFF & $4.48$&$4.50$&$4.46$&$4.49$\\
 & Difference & $0.013$&$0.061$&$0.021$&$0.01$  \\ 

Full  Volume      & Pot  & $6.51$&$6.38$&$6.33$&$6.23$ \\
 & LFF & $6.53$&$6.49$&$6.37$&$6.24$ \\ 
 & Difference&$0.021$&$0.109$&$0.037$&$0.018$
\enddata
\end{deluxetable}

\section{Conclusions}
\label{conclusion}

NOAA 10956 rotated onto the solar disk on 2007 May 15 and on 2007 May 19 the region produced a B9 class flare, starting at 12:34~UT, peaking at 13:02~UT, and ending at 13:19~UT. The flare location suggests that the source of the flare was the loop structure connecting regions 1$^-$ to 2$^+$ in our analysis. According to the \emph{SOHO}/LASCO CME Catalogue\footnote{\url{http://cdaw.gsfc.nasa.gov/CME\_list}} , the region had an associated CME which first appeared at 13:24~UT in the C2 field of view. In addition, a disappearing filament was observed erupting from the lower right section of the regions beginning at 12:31~UT~\citep{long:2008}. 

Our analysis has highlighted a number of physical changes within the magnetic structure of the active region around this time. These are:
\begin{itemize}
\item Magnetic flux cancellation and emergence.
\item Increase in the amount of $\alpha$ or twist along the observed loop.
\item Changes in the connectivity between sub-regions. 
\item Significant changes in the magnetic energy stored in the region.
\end{itemize}

% What do we know
% increase in alpha, unbalanced flux, energy, increase connectivity. 
The increase in the amount of twist or \al{} along the observer loop from 09:35~UT to 11:11~UT, was accompanied by an increase in the magnetic flux within sub-region 1$^+$, and decreases in the magnetic flux within sub-regions 1$^-$ and 2$^+$ and an increase in the amount of magnetic energy stored within the volume. The increase in magnetic energy can be explained by the increase in \al{} used to calculate it. However, as \al{} is obtained from a comparison to EUV observations this increase is taken as real. Following the increase in \al{} at 11:11~UT, the amount of twist along the observed loop decreases to 09:35~UT levels by 14:27~UT. During this time period the magnetic flux in sub-regions 1$^+$, 1$^-$, 2$^+$, and 3$^+$ increases by $0.4-1.5\times10^{20}$~Mx and there is a $1.1\times10^{20}$~Mx decrease in the magnetic flux in region 4$^+$. Additionally, the magnetic energy decreases to 09:35~UT levels over the same time period. Changes in the connectivity from region 1$^-$ can be associated with the changes in \al{} prior to 11:11~UT. However, by 12:47~UT the amount of flux connecting to region 1$^-$ to regions 1$^+$ and 3$^+$ rose significantly. 

From 09:35~UT to 11:11~UT the difference in magnetic energies increased from $1.3\times10^{30}$ to $6.1\times10^{30}$~ergs in the sub-volume centered on the observed loop and from $2.1\times10^{30}$ to $10.9\times10^{30}$~ergs in the full computational domain. This result is particularly important as the energy calculation at 11:11~UT is significantly different from the potential case, as indicated by the error estimate. It should be noted that smaller error bounds may be possible provided that a more detailed analysis of the cost function (smaller $\Delta_{\alpha}$) is performed and that a clear minimum is the cost function exists. By 12:47~UT this difference in magnetic energy in the system had dropped by a factor of three and by 14:27~UT had decreased below the difference at 09:35~UT. The build-up and subsequent decrease in magnetic energy in the system is correlated with the mechanisms thought to be responsible for the release of the B9 flare at 12:30~UT. The method is unable to calculated the free energy of the system due to the limitations of LFF methods~\citep{seehafer:1978}. However, the reported changes in the difference in magnetic energy between the potential and LFF fields provide a estimate for changes in the free energy contained in the region during this time.

The above investigation of magnetic properties of NOAA 10956 has shown that from 09:35~UT there was a significant amount of flux cancellation within the region, accompanied by an increase in the amount of twist and free energy along the observed loop. Following the onset of the flare at 12:34~UT, the amount of twist and magnetic energy in the loops decreased. This was accompanied by the decrease in open flux leaving region 1$^-$ and increase in the magnetic flux in the region. It is suggested that the sudden decrease and subsequent increase in magnetic flux within the active region and the changes in twist and magnetic energy caused small changes in the observed connectivity between sub-regions, Table~\ref{connectivity}. The emergence of magnetic flux within the region and resulting increase in magnetic energy could have resulted in the region passing a critical threshold, passed which it was in an unstable configuration. Under the theory of self-organized criticality this would have resulted in a series of events returning the region to a stable configuration~\citep{vlahos:2004}. Whether the mechanism responsible for passing this critical threshold is the unbalanced emergence of flux within the region or the resulting non-potentiality of the flux ropes remains unknown. The decrease in both \al{} and magnetic energy and increase in connectivity of loop connecting region 1$^-$ to region 3$^+$ suggest that the loop had room to expand and relax into the neighboring sub-region following the flare and eruption in the neighboring sub-region.

%Para on wwieglmann:2005hy we better than other previous work.
The method described in this paper allows for the detailed analysis of the amount of current or $\alpha$ in coronal loop structures and with the added use of the twin perspectives of \emph{STEREO} EUVI images, are an expansion of the methods developed by \citet{wiegelmann:2005} and \citet{carcedo:2003}. These previous studies involved the use of cost functions as a mechanism for constraining the free parameters in extrapolations of the solar corona and were restricted to the investigation of one or two properties and singular instances in time. More advanced NLFF method have been used to investigate the free magnetic energy of active regions~\citep{regnier:2007,Schrijver:2008} and in some instance its evolution~\citep{thalmann:2008,thalmann:2008a}. However these methods have been restricted by the limited time cadence of vector magnetogram data and provide little detail into the physical changes in the structure of active regions surrounding the buildup and release of magnetic energy. While the method developed here are unable to accurately calculate the free energy of the active region, changes in the energy difference can be used as a proxy to changes in the free energy. 

The method has been shown to detect changes in the amount of twist (and hence current) within coronal loop structures using the twin perspective of \emph{STEREO}. Coupling the temporal analysis of the structure of NOAA 10956 with an investigation of the magnetic flux within each sub-region has provided great insight into the evolution and driving forces within the active region. Future work is needed improve the method and remove the need for user input. Expanding the method for the analysis of multiple loop structures in active regions would greatly increase the diagnostic power of the algorithm.

\acknowledgments

We are grateful to the referee for their helpful comments. P. A. Conlon is an IRCSET Government of Ireland Scholar funded under the Irish National Development Plan. The authors would like to thank Dr. R.T.J. McAteer and Dr. D.S. Bloomfield for their helpful discussion and thoughts. The LFF method used in this study is an adaptation of code originally developed by Dr. Thomas R. Metcalf.

\bibliographystyle{apj}
\bibliography{reference} 

\end{document}